\documentclass[letterpaper, 10 pt, conference]{ieeeconf}  

\IEEEoverridecommandlockouts                              

\usepackage[svgnames]{xcolor}
\usepackage{graphicx}            
\usepackage{cite}                
\usepackage{indentfirst}          
\usepackage{latexsym}             
\usepackage{multirow}             
\usepackage{supertabular}         
\usepackage{amsthm}               
\usepackage{mathtools}            
\usepackage{amsfonts}             
\usepackage{url}                  
\usepackage{romannum}             
\usepackage{dsfont}               
\usepackage{rotating}             
\usepackage{gensymb}              
\usepackage{algorithm}            
\usepackage{algpseudocode}        
\usepackage{adjustbox}            
\usepackage{newtxtext,newtxmath}  
\usepackage{booktabs}             
\usepackage{tabularx}             
\usepackage{array}                

\theoremstyle{plain}
\newtheorem{theorem}{Theorem}
\theoremstyle{definition}
\newtheorem{remark}{Remark}
\theoremstyle{plain}
\newtheorem{corollary}{Corollary}
\newtheorem{lemma}{Lemma}

\DeclareMathOperator{\sech}{sech}

\title{\LARGE \bf
Learning Optimal Interaction Weights in Multi-Agents Systems
}

\author{Sara Honarvar and Yancy Diaz-Mercado
\thanks{*This work was supported by the Microsoft Diversity in Robotics and AI Fellowship from the Maryland Robotics Center. }
		\thanks{S. Honarvar, and Y. Diaz-Mercado are with Department of Mechanical Engineering,
			University of Maryland, College Park, MD 20742 
			{\tt\small 
				honarvar@umd.edu, yancy@umd.edu}.}%
\thanks{*This work has been submitted to the IEEE for possible publication. Copyright may be transferred without notice, after which this version may no longer be accessible.}}
    
\begin{document}

\maketitle
\thispagestyle{empty}
\pagestyle{empty}

\begin{abstract}

This paper presents a spatio-temporal inverse optimal control framework for understanding interactions in multi-agent systems (MAS). We employ a graph representation approach and model the dynamics of interactions between agents as state-dependent edge weights in a consensus algorithm, incorporating both spatial and temporal dynamics. Our method learns these edge weights from trajectory observations, such as provided by expert demonstrations, which allows us to capture the complexity of nonlinear, distributed interaction behaviors. We derive necessary and sufficient conditions for the optimality of these interaction weights, explaining how the network topology affects MAS coordination. The proposed method is demonstrated on a multi-agent formation control problem, where we show its effectiveness in recovering the interaction weights and coordination patterns from sample trajectory data. 
\end{abstract}

\section{Introduction}
\begin{figure}[t!]
    \centering
    \includegraphics[width=0.8\linewidth]{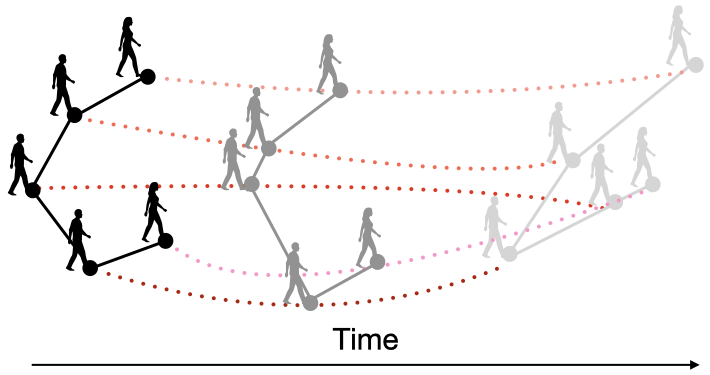}
    \caption{A network of agents in a MAS. The topology of interactions evolves dynamically as agents move and share information, with interaction strengths represented by edge weights. The goal is to learn the edge weight policy from observed trajectories.}
    \label{fig:network of agents}
\end{figure}

Multi-agent systems (MAS) represent collections of autonomous agents working together to accomplish a task, with applications such as crowd navigation \cite{mavrogiannis2023core}, human-robot (swarm) interaction \cite{diaz2017human}, and multi-robot systems \cite{rizk2019cooperative}. Graph representations provide a natural framework to model inter-agent coordination, where nodes represent agents, edges define their interactions, and edge weights quantify the strength of interactions. Understanding these interaction patterns is crucial, as agents must adapt their behaviors based on the environmental context and the states of neighboring agents. The topology of interactions evolves over time as agents move, and communicate (Fig. \ref{fig:network of agents}). Identifying neighborhood (adjacency) relationships between agents and how strongly agents influence each other (i.e., encoded as interaction weights) are essential for understanding the complex network dynamics, such as pedestrian movements in crowds. 

Classical methods such as graph signal processing (GSP) \cite{mateos2019connecting} and optimization-based approaches \cite{ortega2018graph}, primarily focused on identifying graph topology or estimating the Laplacian matrix from data. However, these approaches often assume uniform edge weights across all edges, limiting their ability to capture complex interactions. Recent advances such as self-attention encoders \cite{sebastian2023learning} have sought to recover edge weights from data but generally assume time-invariant weights. While graph neural networks (GNNs) \cite{mohamed2020social, gama2021graph, huang2019stgat} offer increased flexibility in learning interaction patterns, they can be difficult to interpret for performance analysis.

Here, we adopt a control perspective, leveraging weighted consensus algorithms to characterize distributed coordination in MAS \cite{amirkhani2022consensus}.
These algorithms explore how network topology influences performance and convergence, leveraging notions from algebraic graph theory \cite{mesbahi2010graph}.
Weighted consensus provides a foundation for advanced coordination behaviors such as formation control, where constraints like collision avoidance can be enforced to MAS \cite{kamel2020formation}. Desired formation patterns are embedded in the coordination graph through carefully designed edge weights, often utilizing distance-based artificial potential functions (APFs) \cite{mesbahi2010graph}, where the function minimum corresponds to the target formation pattern.
However, designing energy functions for complex topologies remains challenging, as local convexity does not guarantee global convergence in non-convex networks.
%

We propose an inverse optimal control (IOC) framework to learn the interaction behavior directly from data. Existing IOC \cite{zhang2014distributed} and inverse reinforcement learning (IRL) methods aim to recover reward functions that characterize MAS behavior \cite{han2020cooperative}, but often pose restrictive assumptions about system linearity, and time-invariance.

In contrast, this study employs a weighted consensus algorithm to model the interactions between agents, focusing on learning optimal, state-dependent edge weights directly from expert demonstrations via an IOC approach. Unlike previous methods, our framework does not rely on parametric energy functions (i.e., with known functional forms), system linearity, or time-invariance. By deriving optimality conditions, we provide a geometric interpretation of how interaction weights shape agent dynamics, demonstrating the potential to learn coordination behaviors.

The remainder of this paper is organized as follows: Section II outlines key concepts for formulating the MAS coordination problem. In Section III, we formalize the IOC framework, explaining how interaction weights are learned from expert demonstrations and presenting the conditions for optimality. Section IV presents a simulation example, showcasing the effectiveness of the proposed method in recovering edge weights from MAS trajectory data. Finally, Section V concludes the paper with key findings, offers insights for future research, and discusses the broader applicability of the framework to complex, dynamic MAS.
\section{Preliminaries}
\subsection{Notation}

We use $\mathbb{R}$ to denote the set of real numbers, $\mathbb{R}^{d}$ to denote the set of real vectors with $d$ elements, and $\mathbb{R}^{n\times m}$ to represent the set of real $n\times m$ matrices. The vector of all ones in $\mathbb{R}^d$ is denoted by $\mathbf{1}_d$. Let $I_d\in\mathbb{R}^{d\times d}$ represent the $d \times d$ identity matrix, and $\mathds{1}_S(x)$ be the indicator function which returns 1 if $x\in S$ and zero otherwise. When the input to the indicator function is vector valued, we assume the indicator function is applied element wise. For a given set $S$, $|S|$ refers to its cardinality. A weighted, directed graph $G=(\mathcal{V},\mathcal{E},w)$ is a tuple containing a node set $\mathcal{V}$, an ordered edge set $\mathcal{E}$, and a set of weight functions $w$.
The node set, $\mathcal{V} =\{1,\dots,N\}$, identifies each agent. The edge set, $\mathcal{E} \subset \mathcal{V} \times \mathcal{V} $, encodes the topology of interaction. The weight functions $w : \mathcal{V} \times \mathcal{V} \mapsto \mathbb{R}$, assumed to be unknown, represent how strongly agents influence their neighbors. The matrices $D^{in}, D^{out}\in\mathbb{R}^{|\mathcal{V}|\times|\mathcal{E}|}$ refer to in-degree and out-degree matrices of a given graph, which are defined later in the text. The square diagonal matrix formed from the elements in a vector $a$ is written as $\text{diag}(a)$. For any vector $p$, $[p]_k$ represents its $k$-th element, and $p^T$ refers to its transpose. For a matrix $P$, $[P]_{ik}$ is the entry in the $i$-th row and $k$-th column. The Euclidean distance ($\ell^2$ norm)  between two vectors $p$ and $q$ is $\|p - q\|$. The Kronecker product is represented by $\otimes$, and the Hadamard (element-wise) product by $\odot$. 
\begin{lemma}\label{hadamard properties-original}
The Hadamard product (i.e., element-wise multiplication) of two compatible vectors $a\in\mathbb{R}^d$ and $b\in\mathbb{R}^d$ can be expressed as the matrix multiplication of the corresponding diagonal matrix of one vector by the other vector \cite{liu2008hadamard}:
\begin{equation}
    a \odot b = \text{diag}(a)  b = \text{diag}(b) a,
\end{equation}
\begin{equation}
    \text{diag}(a) \text{diag}(b) = \text{diag} (a \odot b),
\end{equation}
\begin{equation}\label{diag}
    \text{diag}(a) = a \mathbf{1}^T_d \odot I_d,
\end{equation}
where $\mathbf{1}_d$ is a vector of all ones and $I_d$ is the identity matrix.
\end{lemma}
The hat notation ($\hat{\square}$) refers to the observed values, and an $o$ subscript ($\square_o$) denotes nominal or reference values. The set of continuously differentiable functions is $\mathcal{C}^1$, and $\mathcal{L}^1([\delta,\Delta])$ denotes the set of Lebesgue-integrable functions over $[\delta,\Delta]$. The weight matrix $\mathbb{W}(\alpha(x), u(s))$ is a diagonal matrix, whose diagonal elements are dependent on the control variable $u(s)$ and a function of the agent state $\alpha(x)\in\mathbb{R}^{|\mathcal{E}|}$, where $[\alpha(x)]_{k} = \alpha_k(x) = \|x_i-x_j\|$ represents the Euclidean distance between agents (nodes) $i$ and $j$ at edge $k = (j,i) \in \mathcal{E}$. 
\subsection{Multi-Agent Dynamics and Interaction Representation}
Consider a system of $N$ agents, where each agent's state at time $t$ is represented by a vector, $x_i(t)\in \mathbb{R}^d$. The overall system state is denoted by the ensemble state vector $x(t) = [x_1^T(t), \dots, x_N^T(t)]^T \in \mathbb{R}^{d\times N}$. The interactions among agents are modeled as a directed graph $G = (\mathcal{V},\mathcal{E},w)$, where $N=|\mathcal{V}|$ is the number of nodes, and $m=|\mathcal{E}|$ is the number of edges.
For each edge $k = (j,i) \in \mathcal{E}$, the interaction between agents $i$ and $j$ is encoded as an unknown weight $w_k\left(\alpha_k(x), u(s)\right)$, which depends on the distance between agents, $\alpha_k(x)$, and a control variable $u(s)$. The control variable (policy), $u(s)$ is a non-parametric, scalar-valued function for any $0\leq\delta\leq s\leq \Delta <\infty$. We assume that $u(s)\in \mathcal{L}^1([\delta,\Delta])$.

Here, as in \cite{SaraModel}, we assume that interaction dynamics are decouple from an individual agent's main planner. Given the edge set, the goal is to recover the state-dependent edge weights, $w_k\left(\alpha_k(x), u(s)\right)$, which best explain the observed coordination behaviors. The system dynamics are thus given at the node-level for all $i\in \mathcal{V}$ by
\begin{equation}:
    \dot{x}_i = h_i(x_i) + \!\!\!\!\!\!\!\!\sum_{\{k\in\mathcal{E} \, | \, k \, =  \, (j,i)\}} \!\!\!\!\!\!\!\! w_{k}(\alpha_k(x),u(s))(x_j(t)-x_i(t))
\end{equation}
or at the ensemble-level as: 
\begin{equation}
    \dot{x} = h\left(x\right)-\Big(L^{in}\big(w\left(\alpha(x), u(s)\right)\big)\otimes I_d \Big)x,
\end{equation}
The dynamics consists of two terms: the first term, $h\left(x\right) \in \mathbb{R}^{dN}$, refers to the goal reaching dynamics (i.e., ``goal control") and the second term captures agent interaction through weighted consensus dynamics (i.e., ``agreement control"). The in-Laplacian matrix is given by: $$ L^{in}\big(w\left(\alpha(x), u(s)\right)\big) = D^{in}\mathbb{W}\left(\alpha(x),u(s)\right)D^T,$$
where $D = D^{in}-D^{out} \in \mathbb{R}^{N \times m}$ is the incidence matrix, given as follows:
\begin{equation}
  [D]_{ik}=\begin{cases}
      1 & \textit{if } \exists k = (j,i)\in\mathcal{E}\\
      -1 & \textit{if } \exists k = (i,j)\in\mathcal{E}\\
      0 & \text{otherwise}
  \end{cases}  
\end{equation} 
with $D^{in}$, and $D^{out}$ as: 
\begin{equation}
\begin{gathered}
    [D]^{in}_{ik}= \begin{cases} 
        1 & \text{if } \exists k = (j, i)\in\mathcal{E} \\ 
        0 & \text{otherwise}, 
        \end{cases}\\
            [D]^{out}_{ik}= \begin{cases} 
        1 & \text{if } \exists k = (i, j)\in\mathcal{E} \\ 
        0 & \text{otherwise} 
    \end{cases}
\end{gathered}
\end{equation}
The diagonal weight matrix is given by $\mathbb{W}\left(\alpha(x),u(s)\right) = \text{diag}\left(w\left(\alpha(x), u(s)\right)\right)\in \mathbb{R}^{m\times m}$.


For the weights to be smooth, we parameterize them as follows:
\begin{equation}\label{weight by u}
[w(\alpha(x), u(s))]_k = \int_{\delta}^{\alpha_k(x)} u(s) \, ds = \int_{\delta}^{\Delta} u(s)\mathds{1}_{{s<\alpha_k(x)}}\, ds,
\end{equation}
where $\mathds{1}_{{s<\alpha_k(x)}}$ is the indicator function, defined as\[
[\mathds{1}_{{s<\alpha(x)}}]_k=\begin{cases}
    1 & \textit{if } s<\alpha_k(x),\\
    0 & \text{otherwise}.
\end{cases}\]

Using the weight representation, we rewrite the node-level agent dynamics as
\begin{equation}\label{node-level dynamics}
    \dot{x}_i =\int_{\delta}^{\Delta} \frac{1}{\Delta-\delta} h_i(x_i)- \!\!\!\!\!\!\!\!\sum_{\{k\in\mathcal{E} \, | \, k \, =  \, (j,i)\}} \!\!\!\!\!\!\!\!u(s) \mathds{1}_{s<\alpha_k(x)} (x_i-x_j)\, ds. 
\end{equation}
Dynamics of this form has been shown to have applications in various settings, including human interaction models \cite{SaraModel} and flocking models like the Cucker-Smale model \cite{cucker2007emergent}. 

We can similarly rewrite the dynamics at the ensemble-level:
\begin{equation}\label{ensemble dynamics}
    \dot{x} = \int_{\delta}^{\Delta} \Big[ \frac{1}{\Delta-\delta} h(x) - \mathcal{D}^{in}\Big( \text{diag}\big(u(s) \mathds{1}_{s<\alpha(x)}\big)\otimes I_d \Big) \mathcal{D}^T x\,\Big] ds,
\end{equation}
where $\mathcal{D}^{in}=D^{in}\otimes I_d$, and $\mathcal{D}=D\otimes I_d$. 

\begin{lemma}\label{hadamard properties-results}
The diagonal term in \eqref{ensemble dynamics} can be interchanged as:
\begin{align}
 \mathcal{D}^{in} \Big([\text{diag}\big(u(s) \mathds{1}_{s<\alpha(x)}\big)]\otimes I_d \Big) \mathcal{D}^T x = \nonumber\\\mathcal{D}^{in}  \Big([\text{diag}(\mathcal{D}^T x) u(s) \mathds{1}_{s<\alpha(x)} ]\otimes \mathbf{1}_d\Big).
\end{align}
\proof
    Follows from the Hadamard properties in Lemma \ref{hadamard properties-original}.   
    Using the properties of Hadamard product from \eqref{diag}, the diagonal term can be equivalently represented as
    \begin{equation}
     \begin{gathered}
    [\text{diag}\big(u(s) \mathds{1}_{s<\alpha(x)}\big)] = [ u(s) \mathds{1}_{s<\alpha(x)} \mathbf{1}_m^T\odot  I_m],
    \end{gathered}       
    \end{equation}
    Therefore, by exploiting the Hadamard properties once more, we can interchange the diagonals as
    \begin{equation}
     \begin{gathered}
        \mathcal{D}^{in}  \Big([\text{diag}\big(u(s) \mathds{1}_{s<\alpha(x)}\big)]\otimes I_d \Big) \mathcal{D}^T x = \\ \mathcal{D}^{in}  \Big([u(s) \mathds{1}_{s<\alpha(x)}\mathbf{1}_m^T\odot  I_m] \otimes I_d \Big) \mathcal{D}^T x =\\
        \mathcal{D}^{in}  \Big([\text{diag}(\mathcal{D}^T x) u(s) \mathds{1}_{s<\alpha(x)} ]\otimes \mathbf{1}_d\Big).
    \end{gathered}       
    \end{equation}
\endproof
\end{lemma}
\begin{corollary}
For the special case of $d=1$, the interaction dynamics simplify to: 
\begin{align}
D^{in}  \text{diag}(u(s) \mathds{1}_{s<\alpha(x)})  D^Tx =\nonumber D^{in}  \text{diag}\big(D^T x\big)  u(s) \mathds{1}_{s<\alpha(x)}.
\end{align}  
\end{corollary}
According to Lemma \ref{hadamard properties-results}, the integrand in \eqref{ensemble dynamics} becomes:
\begin{multline}\label{another versian of dynamics}
  \Tilde{f}(x,u(s)) =\\ \frac{1}{\Delta-\delta} h(x)- \mathcal{D}^{in}  \Big([\text{diag}(\mathcal{D}^T x) u(s) \mathds{1}_{s<\alpha(x)} ]\otimes \mathbf{1}_d\Big).  
\end{multline}
\section{Inverse Optimal Control Problem}
The objective of this work is to learn the optimal interaction weights that govern MAS coordination by observing expert demonstrations. We formulate this as an IOC problem, aiming to recover the state-dependent edge weights, $w\left(\alpha(x), u(s)\right)$ as characterized by $u(s)$, that best explain the observed MAS behavior. Note that the decision variable, $u(s)$, is not a function of time. Instead, it is defined over all possible pair-wise interaction distances.

In classical Bolza OC problems \cite{liberzon2011calculus}, the system seeks to minimize a cost function. Here, we learn this cost indirectly from observed trajectories to derive the control policies without explicitly defining the performance index. Given a set of observed trajectories, $\hat{x}(t) = [\hat{x}_1^T,...,\hat{x}_N^T(t)]^T$, we define the IOC problem for a continuous-time, nonlinear dynamical system as:
\begin{equation}\label{Optimal Control Problem}
\begin{aligned}
\min_{u(s)} J(x,u)&= \min_{u(s)} \int_{t_0}^{t_f} \int_{\delta}^{\Delta} 
\underbrace{Q\left(x(\tau),\hat{x}(\tau)\right)}_{\text{Temporal term}}+\underbrace{G\left(u(s)\right)}_{\text{Spatial term}} ds d\tau \\&+\psi\left(x(t_f),\hat{x}(t_f)\right)\\
\text{s.t.} \quad \dot{x} &=h\left(x(t)\right)-\Big(L^{in}\left(w\left(\alpha(x), u(s)\right)\right)\otimes I_d \Big)x,\\
&\text{Given } x(t_0)=\hat{x}(t_0) \\
\end{aligned}
\end{equation}

The temporal term, $Q\left(x(\tau),\hat{x}(\tau)\right)$ in the cost penalizes for deviations between the estimated and true trajectories over time. The spatial component, $G\left(u(s)\right)$ is a regularization term on the control policy. The final cost, $\psi(x(t_f),\hat{x}(t_f))$, penalizes deviations between the estimated and true trajectories at the final time, $t_f$. 
For conciseness, we express the dynamics as $\dot{x} = f\left(x, u(s)\right)$, where:
 $$ f\left(x, u(s)\right) = h(x)- \Big(L^{in}\left(w\left(\alpha(x),u(s)\right)\right) \otimes I_d \Big) x,$$ 
which is Lipschitz with $\alpha(x) \in \mathcal{C}^1$.

The aim is to solve for the optimal policy $u^*(s)$ that minimizes the cost $J(x,u(s))$ and recovers the optimal interaction weights, $w^*(\alpha(x^*), u^*(s))$.
\subsection{Learning Multi-agent Interaction Dynamics}
We aim to learn optimal MAS interaction by solving the IOC problem in \eqref{Optimal Control Problem} to recover the interaction edge weights.
\subsection{Optimality Conditions}
\begin{theorem}\label{optimality condition}
Given a network topology of interaction and the regularization cost,
\begin{equation}\label{regularizer G(u(s))}
    G(u(s))=\frac{1}{2}\frac{\|u(s)-u_o(s)\|^2}{t_f-t_0},
\end{equation}
for an arbitrary nominal policy $u_o(s)\in \mathcal{L}^1([\delta,\Delta])$, the optimal signal $u^*(s)$ satisfies the following optimality condition:
\begin{equation}
\begin{aligned}
& u^*(s)=u_o(s)\\&- \frac{1}{t_f-t_0}\int_{t_0}^{t_f} \sum_{i=1}^N\sum_{(j,i)\in \color{red}\mathcal{E}}\color{black}\lambda_i^{*^T} (x_i^*(\tau)-x_j^*(\tau))\textbf{1}_{s<\|x_i^*-x_j^*\|} \, d\tau,
\end{aligned}
\end{equation}
which explicitly accounts for the network topology, as encoded by edge set $\color{red}\mathcal{E}$, and where $x_i^*$ and $\lambda_i^*$ are the optimal state and co-states for $i=1,...,N$. The learned control $u^*(s)$ minimizes the cost, resulting in the optimal interaction weights as:
$$
w^*(\|x_i^*-x_j^*\|, u^*) = \int_{\delta}^{\|x_i^*-x_j^*\|} u^*(s) \, ds, \forall (j,i) \in \mathcal{E}.
$$
\end{theorem}
\proof
    We use a calculus of variation argument.
    Let 
    \begin{equation}\label{definition of I cal}
    \mathcal{I}\left(x(\tau),u(s)\right) =Q(x(\tau),\hat{x}(\tau))+G(u(s)),    
    \end{equation}
    To derive the optimality conditions, we first construct the augmented cost as
    \begin{equation}\label{cost}
        \begin{aligned}
      &\Tilde{J}(x,\dot{x},u(s)) =  \psi\left(x(t_f)\right)\\
      &+\int_{t_0}^{t_f} 
\int_{\delta}^{\Delta}\Big[\mathcal{I}\left(x(\tau),u(s)\right) + \lambda^T (\frac{\dot{x}}{\Delta-\delta}-\Tilde{f}(x,u(s)))\Big]\  ds d\tau      
        \end{aligned}
    \end{equation}
    where $\lambda \in R^{d\times N}$ is the co-state vector enforcing the dynamic constraints. Let $u(s) \rightarrowtail u(s)+\varepsilon \mu(s)$, and consequently, due to the Lipschitz dynamics, $x(t)\rightarrowtail x(t)+\varepsilon \eta(t)$, and $\dot{x}(t)\rightarrowtail \dot{x}(t)+\varepsilon \dot{\eta}(t)$. Using Taylor series expansion, we construct the Gateaux derivative \cite{liberzon2011calculus} of the cost $\delta \Tilde{J}$ as follows 
    \begin{equation}\label{variation of augmented cost}
        \begin{aligned}
         \delta \Tilde{J} &=\lim_{\varepsilon \to 0}\tfrac{1}{\varepsilon}\left(\tilde{J}(u+ \varepsilon \mu(s))-\tilde{J}(u(s))\right)\\&=\frac{\partial \psi}{\partial x}\bigg|_{x=x(t_f)}\eta(t_f)+\lambda^T(t_f)\eta(t_f)\\
    &+\displaystyle\int_{t_0}^{t_f} \Big[-\dot{\lambda}^T+\int_{\delta}^{\Delta}\Big(\frac{\partial\mathcal{I}}{\partial x} \bigg|_{x=x}
     -\lambda^T \frac{\partial \Tilde{f}}{\partial x}\Big)ds\Big]\eta(\tau)\, d\tau\\&+\int_{t_0}^{t_f} \int_{\delta}^{\Delta}\Big(\frac{\partial\mathcal{I}}{\partial u(s)}  - \lambda^T\frac{\partial \Tilde{f}}{\partial u(s)} \Big)\mu(s) \, ds d\tau,
        \end{aligned}
    \end{equation}
    We then set $\delta \Tilde{J} = 0$ to find the necessary conditions for optimality. The details of how to construct the Gateaux derivative and derive the necessary conditions for optimality are given in appendix A.
    
    First-order necessary condition for optimality is given by the following co-state equation:
    \begin{multline}\label{co-state}
            \dot{\lambda}^T  = \int_{\delta}^{\Delta} \Big( \frac{\partial\mathcal{I}}{\partial x}  - \lambda^T \frac{\partial \Tilde{f}}{\partial x} \Big)\, ds 
            =\frac{\partial Q}{\partial x}-\lambda^T \frac{\partial h(x)}{\partial x}\\
            +\lambda^T\mathcal{D}^{in} \mathbb{W}  \mathcal{D}^T + \lambda^T\mathcal{D}^{in} \odot x^T\mathcal{D}  \big(\text{diag}\left(u(\alpha(x)\right) \otimes \mathbf{1}_d\big)\frac{\partial \alpha(x)}{\partial x}.
    \end{multline}
    Note that $\text{diag}(\mathcal{D}^Tx)=\text{diag}(x^T\mathcal{D})$ and the $kj$-th block of $\left[\frac{\partial \alpha}{\partial x}\right]$ is (refer to appendix A):
    \begin{equation}
    \left[\frac{\partial \alpha}{\partial x}\right]_{kj}  =  \frac{\partial \alpha_k}{\partial x_j}  =
         \begin{cases}
          \frac{(x_i-x_j)^T}{\|x_i-x_j\|} & \textit{if } \exists k \in (j,i)\\
          -\frac{(x_i-x_j)^T}{\|x_i-x_j\|} & \textit{if } \exists k \in (i,j)\\
          0^T & \textit{o.w}.
     \end{cases}
    \end{equation}
    The boundary condition is then given by:
    \begin{equation}
     \lambda (t_f)=-\Big(\frac{\partial \psi}{\partial x}\Big)^T\bigg|_{x=x(t_f)}.
    \end{equation}
    
    We have an additional integral constraint, giving us the optimal policy
    \begin{equation}
        \begin{aligned}      \Tilde{J}_u:=\int_{t_0}^{t_f}\frac{\partial\mathcal{I}}{\partial u}  - \lambda^T\frac{\partial \Tilde{f}}{\partial u(s)}   \,d\tau \overset{!}{=}0,
        \end{aligned}
    \end{equation}
    Expanding, we get
    \begin{multline}\label{cost Ju}
    \Tilde{J}_u= u-u_o+ \tfrac{1}{t_f-t_0}\int_{t_0}^{t_f}\lambda^T \mathcal{D}^{in}\text{diag}(\mathcal{D}^Tx)[\mathds{1}_{s<\alpha(x)}\otimes \mathbf{1}_d] \, d\tau
         \\
         = u-u_o+ \tfrac{1}{t_f-t_0}\int_{t_0}^{t_f} \sum_{i=1}^N\sum_{j\in \mathcal{N}_i^{in}}\lambda_i^T (x_i-x_j)\mathds{1}_{s<\|x_i-x_j\|} \, d\tau
    \end{multline}
    By setting \eqref{cost Ju} to zero, we can obtain the optimal policy, $u^*(s)$, and consequently the optimal interaction weights.
\endproof

\begin{corollary}
The IOC problem defined in \eqref{Optimal Control Problem} satisfies the second-order sufficient condition for optimality, ensuring the solution corresponds to a strict minimum of the cost functional.
\proof
    The Hamiltonian is constructed as
    $$ H = \int_\delta^\Delta \mathcal{I}(x,u(s))+\lambda^T\tilde{f}(x,u(s))\, ds,$$ where $\mathcal{I}(x,u(s))$ is the integrand cost based on \eqref{definition of I cal}, and $\tilde{f}(x,u(s))$ is given by \eqref{another versian of dynamics}. The Hessian $\frac{\partial^2{H}}{\partial{u(s)^2}} > 0$ for all values of $s$ (refer to Appendix B for more details). Therefore, the second-order sufficient condition for optimality is satisfied as per \cite{liberzon2011calculus}.
\endproof
\end{corollary}
\begin{remark}
    No assumptions are made on the nominal policy, $u_o(s)$, in our approach. The optimal policy, $u^*(s)$, converges to $u_o(s)$ outside the interval $[\delta, \Delta].$
\end{remark}
\subsection{Learning Interaction Weight Algorithm}
Based on Theorem \ref{optimality condition}, the optimal weights are derived by leveraging the topology of interaction. The first-order optimality condition on $u(s)$ can serve as a gradient to learn the optimal policy via iterative approaches such as gradient descent. Algorithm 1 outlines the steps to learn the interaction weights, $w^*$, using gradient descent with an Armijo line search \cite{armijo1966minimization}.
\begin{algorithm}[b!]
\label{Interaction algorithm}
\caption{Interaction Weights Learning}
\begin{algorithmic}
\State Initialize the initial state $x_0$ and the initial control $u(s)=u_0$
\State Set the iteration counter: $\text{iter} \gets 0$
\While {$\text{iter} < \text{Maximum Iteration}$}
\State \textbf{Forward Pass:}
\For{$t=t_0:t_f$}
\State Propagate the state forward as follows:
\State \quad\quad $\dot{x}(t) = f(x(t)) - L^{\text{in}}\left(x(t),u(s)\right)x(t)$
\EndFor
\State Calculate the co-state at the terminal time:
\State \quad\quad $\lambda(t_f) =  -\Big(\frac{\partial \psi}{\partial x}\Big)^T\bigg|_{x=x(t_f)}$
\State \textbf{Backward Pass:}
\For{$t=t_f:t_0$}
\State Solve the co-state equation (\ref{co-state}) backwards in time \State from terminal conditions
\EndFor
\If{$\left|\ell\tilde{J}_u(s)\right| < \epsilon$}
\State The gradient norm is sufficiently small; terminate
\Else
\State Use the \textbf{Armijo Line Search} \cite{armijo1966minimization} to find the \State optimal step size, $\ell$
\EndIf
\State Update $u$ using gradient descent:
\State $u(s) = u(s) - \ell \tilde{J}_u$
\State Increment the iteration counter: $\text{iter} \gets \text{iter}+1$
\EndWhile
\end{algorithmic}
\end{algorithm}
\section{Simulation Example}
To validate the proposed IOC approach, we perform a simulation case study to evaluate its effectiveness in recovering the true edge weights from MAS trajectory demonstrations.

\subsection{Generating Demonstrations}
We start by creating an example of decentralized coordination among agents that are tasked with reaching their goals while maintaining a specific formation pattern. The resulting trajectories provide the ground truth for the IOC problem, which aims to solve for the interaction weights.

\subsubsection{Case Study}
In our example, we consider a set of $N=8$ agents, connected by the edge set $\mathcal{E}=\{(1,3),(2,5),(3,8),(4,5),(4,6),(6,7),(7,8)\}$. We use random initial and goal locations for each of these agents.
The dynamics combine goal-directed behavior, modeled using an exponential motion planner, with agreement control terms. 
\begin{itemize}
    \item \textbf{Goal Control:}
    We use a soft-saturation proportional regulator as our motion planner to lead the agents toward their goal location, $x_{i,g}$. Let $\Tilde{x}_i=x_{i,g}-x_i$. The motion planner is given by:
        \begin{equation*}
        [h(\tilde{x})]_i=h_i(\tilde{x}_i)=\begin{cases} k\tanh{\left(\frac{\|\tilde{x}_i\|}{k}\right)}\frac{\tilde{x}_i}{\|\tilde{x}_i\|},& \|\tilde{x}_i\|^3<\epsilon \\ 0 & \text{otherwise}\end{cases}
        \end{equation*}
        where $\tanh{(x)} = \frac{e^{2x}-1}{e^{2x}+1}$ is the hyperbolic tangent, and $k$ is a scalar gain.
        
        The gradient $\frac{\partial h}{\partial x}$ is a block diagonal matrix, and the $i$th block is $$\textstyle\frac{\partial h_i}{\partial x_i}= \sech^2\left(\frac{\|\tilde{x}_i\|}{k}\right)\frac{\Tilde{x}_i\Tilde{x}_i^T}{\|\Tilde{x}_i\|^2}+k\tanh{\left(\frac{\|\tilde{x}_i\|}{k}\right)}\Big[\frac{I}{\|\Tilde{x}_i\|}-\frac{\Tilde{x}_i\Tilde{x}_i^T}{\|\Tilde{x}_i\|^3}\Big]$$ if $\|\Tilde{x}_i\|^3<\epsilon$ and $-I_d$, otherwise,
        where $\sech(x)$ is the hyperbolic secant.
        \begin{remark}
            This study primarily focuses on learning interaction behaviors; therefore, the design of the goal control behavior considered is arbitrary.
        \end{remark}
\end{itemize}
\begin{itemize}
    \item \textbf{Agreement Control:}
    We consider a straight line for our nominal policy: $u_o(s) = 3(s-d)$. While the true policy $\hat{u}(s) = 3(s-d)^2$, has a quadratic form, where $d=0.3$ is the desired separation between agents, and $s\in[\delta,\Delta]$, with $\delta=0.15$ and $\Delta=3.02$. The edge weights for the agreement control in the ground truth, which our proposed algorithm aims to recover, are given by \eqref{weight by u}:
    \begin{equation}
    \hat{w}_{ij}(\|\hat{x}_i(t)-\hat{x}_j(t)\|) = \left(\|\hat{x}_i(t)-\hat{x}_j(t)\| - d\right)^3.
    \end{equation}
\end{itemize}
    \subsection{Reconstructing Interaction Weights}
    We apply the proposed IOC approach (Algorithm 1) to reconstruct edge weights that best explain the demonstrated coordination patterns. We consider the following temporal cost, $Q$, and final cost, $\psi$ for our demonstration 
    \begin{equation}
        \begin{gathered}
    Q(x(\tau),\hat{x}(\tau))=\frac{1}{2}\frac{\|x(\tau)-\hat{x}(\tau)\|^2}{\Delta-\delta},\\
    \psi(x(t_f),\hat{x}(t_f))=\frac{1}{2}\|x(t_f)-\hat{x}(t_f)\|^2.       
        \end{gathered}
    \end{equation}
    
\subsection{Simulation Results}
\begin{figure}[t!]
    \centering
        \centering \includegraphics[width=\linewidth]{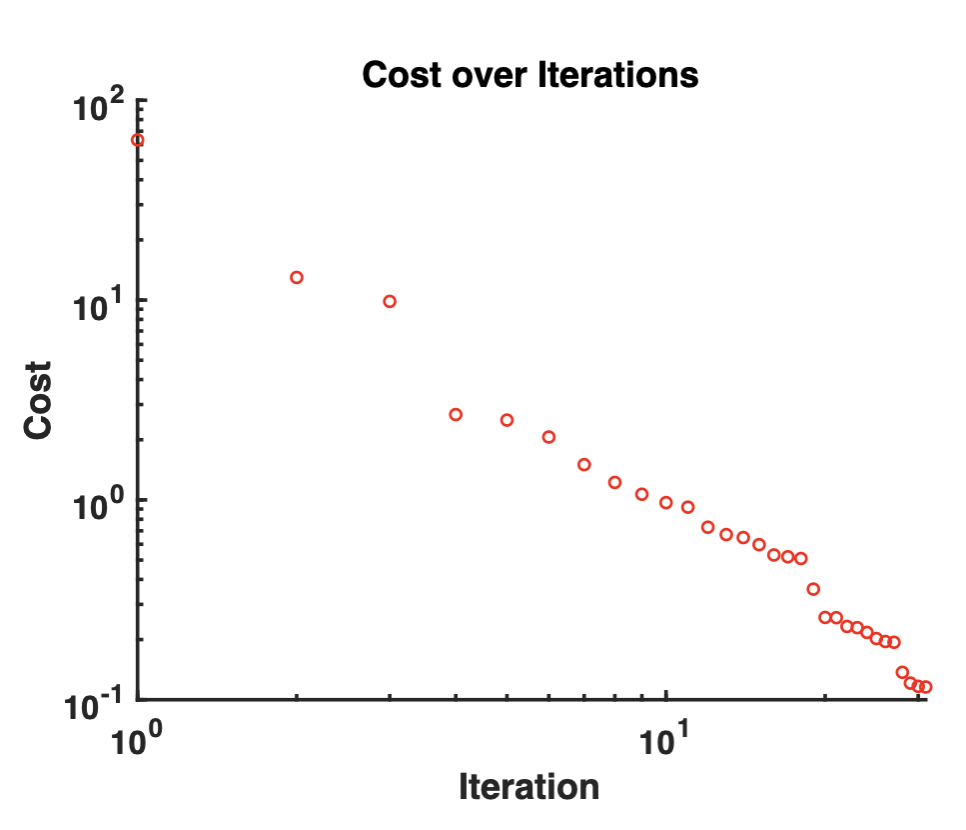}
        \caption{Total cost across iterations}
        \label{fig:total_cost}
\end{figure}

\begin{figure}[t!]
        \centering         \includegraphics[width=1\linewidth]{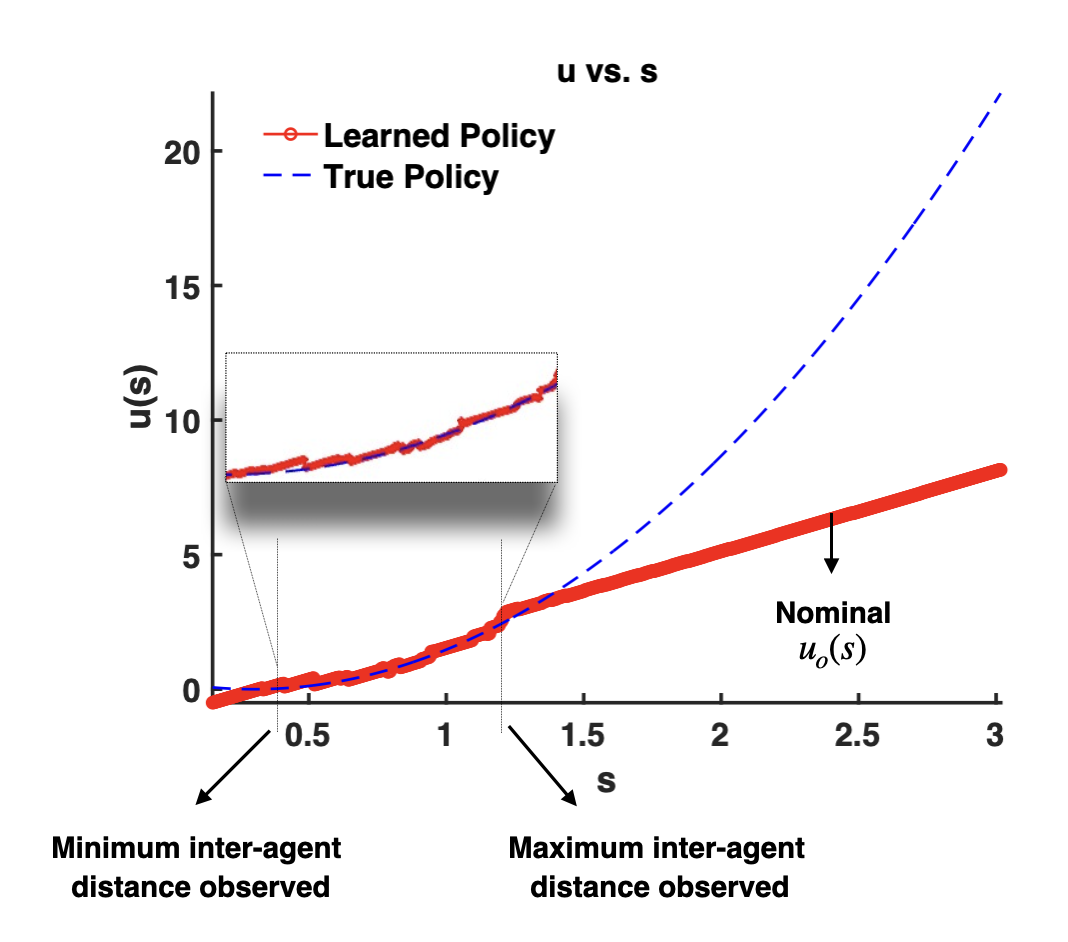}
        \caption{Comparing learned (red) vs. true (blue) policy, $u(s)$, over $s$}
        \label{fig:us_variation}
\end{figure}

\begin{figure}[t!]
    \centering
 \includegraphics[width=\linewidth]{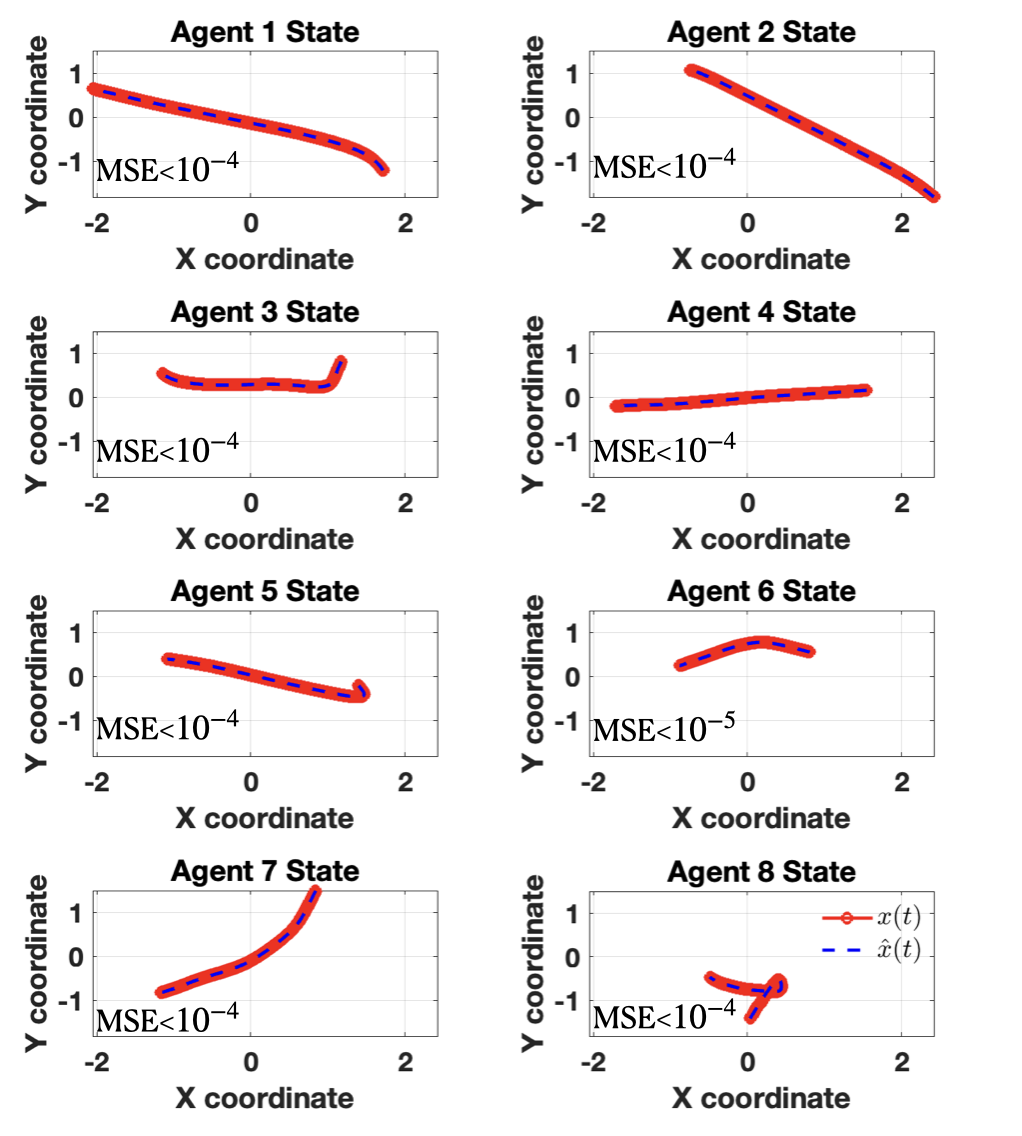}
    \caption{Comparing the simulated trajectories (red) vs. the ground truth trajectories (blue}
    \label{fig:trajectories}

    \centering   \includegraphics[width=1\linewidth]{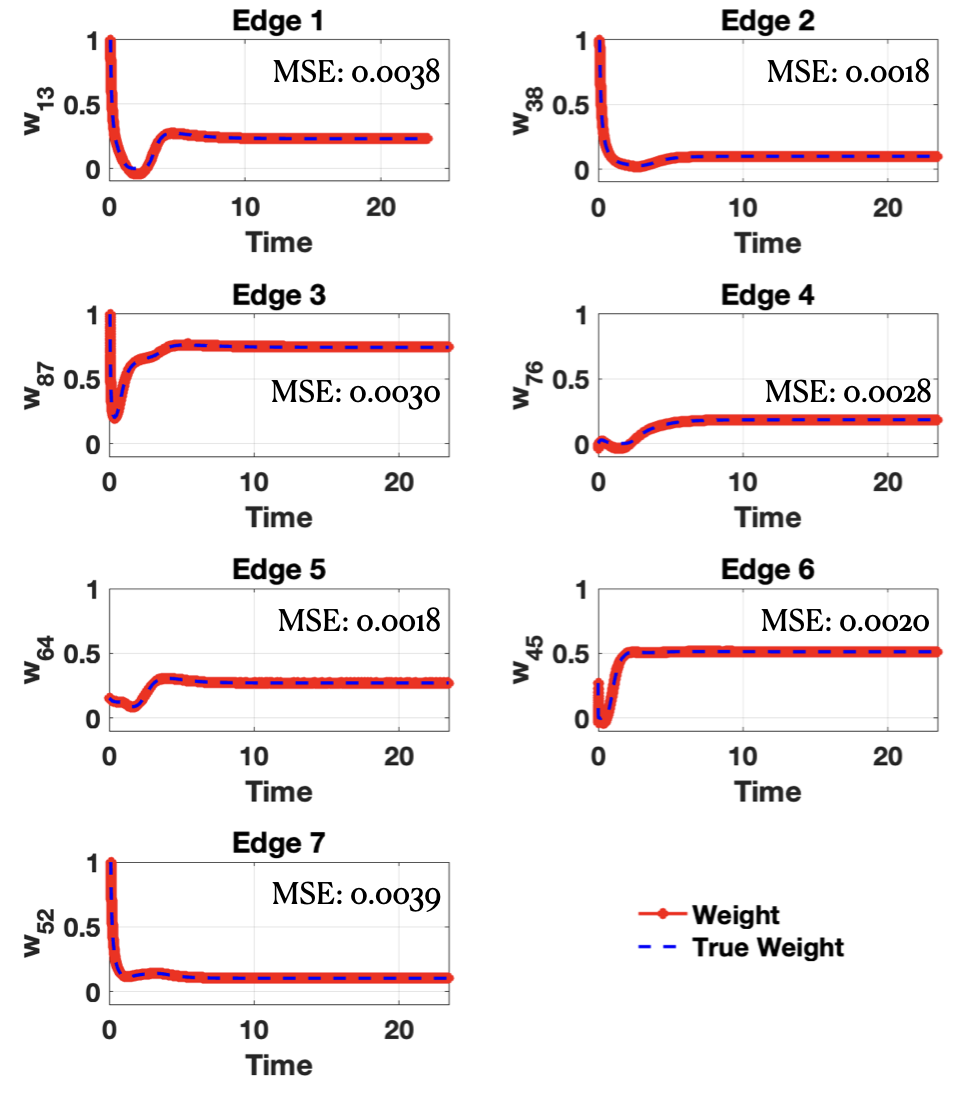}
    \caption{Comparing the learned interaction weights for each edge (red) vs. the ground truth edge weights (blue)}
    \label{fig:weights}
\end{figure}

As seen in Fig. \ref{fig:total_cost}, the algorithm converges quickly. Fig. \ref{fig:us_variation} shows how the policy, $u(s)$, is recovered as a function of the distance between agents. The learned policy, matches the true policy within the range of inter-agent distances observed in the data, while outside this range, it converges to the nominal $u_o$, which for this example is assumed to be a straight line. 
Fig. \ref{fig:trajectories} and Fig. \ref{fig:weights} demonstrate that the algorithm accurately recovers the MAS trajectories and evolution of the edge weights over time, respectively, as evidenced by the low mean squared error (MSE) between the simulation and experimental results.

\section{Conclusion and Future Work}
This paper presented an IOC framework for learning dynamic, state-dependent interaction weights in MAS using consensus algorithms. By observing coordination patterns, our approach provides a geometric and analytic understanding of how these weights influence MAS behavior, without assuming system linearity and time-invariance, making it suitable for complex, nonlinear scenarios. 

The optimality conditions reveal a geometric link between system states and co-states, offering insights into how interaction topology impacts optimality and algorithm convergence. Future work will explore these geometric properties to design topologies that enhance convergence and extend the framework to directed graphs with time-varying topology and heterogeneous edge weights. These insights could improve network structure design and enhance GNN inference.




\section*{Appendix}
In this section, we provide the necessary details for deriving the optimality conditions:
\subsection{Derivation of Optimality Conditions}
Let's consider the general case of setting
\begin{equation}\label{definition of I}
\mathcal{I}\left(x(\tau),u(s)\right) =\frac{1}{2}\frac{\|x(\tau)-\hat{x}(\tau)\|^2}{\Delta-\delta}+\frac{1}{2}\frac{\|u(s)-u_o(s)\|^2}{t_f-t_0}.    
\end{equation}
In order to construct the optimality conditions, we first construct the augmented cost, $\Tilde{J}(x,\dot{x},u(s))$, as in \eqref{cost}.

Let $u(s) \rightarrowtail u(s)+\epsilon \mu(s)$ and consequently, due to Lipschitz dynamics, $x(t)\rightarrowtail x(t)+\epsilon \eta(t)$, and $\dot{x}(t)\rightarrowtail \dot{x}(t)+\epsilon \dot{\eta}(t)$. Using the calculus of variation, we need to set $\delta \Tilde{J} \overset{!}{=} 0$ to find the optimal policy and first-order necessary conditions for optimality, where $\delta \Tilde{J}$ is given by the Gateaux derivative \cite{liberzon2011calculus} as:
\begin{multline}\label{augmented cost variation 2}
\delta \Tilde{J} = \lim_{\epsilon \rightarrow 0} \frac{1}{\epsilon}\left[\Tilde{J}\left(x + \epsilon \eta, \dot{x} + \epsilon \dot{\eta}, u + \epsilon \mu(s)\right) - \Tilde{J}(x, \dot{x}, u(s))\right] \\
= \lim_{\epsilon \rightarrow 0} \frac{1}{\epsilon}\Bigg[\psi\left(x(t_f) + \epsilon \eta(t_f)\right) - \psi\left(x(t_f)\right)\\
+\int_{t_0}^{t_f} \int_{\delta}^{\Delta} I\left(x(\tau) + \epsilon \eta(\tau),u(s)+\epsilon \mu (S)\right) -\mathcal{I}\left(x(\tau),u(s)\right)\\
+\lambda^T\Big(\frac{1}{\Delta-\delta}\left({\dot{x}} + \epsilon \dot{\eta}(t) - {\dot{x}}\right)\\
- \Tilde{f}\left(x + \epsilon \eta,  u + \epsilon \mu(s)\right) 
 + \Tilde{f}\left(x, u \right)\Big) \, ds d\tau
\Bigg]
\end{multline}

The Taylor expansion of $I(x+\epsilon \eta, u+\epsilon \mu(s))$ around $(x,u(s))$ is given by
\begin{multline}
\mathcal{I}(x+\epsilon \eta, u(s)+\epsilon \mu(s)) = \\\mathcal{I}(x, u(s)) + \epsilon \frac{\partial\mathcal{I}}{\partial x} \eta  + \epsilon \frac{\partial\mathcal{I}} {\partial u(s)} \mu(s),
\end{multline}
Similarly, the Taylor expansion of $\Tilde{f}(x+\epsilon \eta, u+\epsilon \mu(s))$ around $(x,u(s))$ is given by
\begin{multline}
\Tilde{f}(x+\epsilon \eta, u+\epsilon \mu(s)) =\\ \Tilde{f}(x, u(s)) + \epsilon \frac{\partial \Tilde{f}}{\partial x} \eta  + \epsilon \frac{\partial \Tilde{f}} {\partial u(s)} \mu(s),
\end{multline}
where using lemma \ref{hadamard properties-results}, 

\begin{multline*}\label{gradient of ftilde for d>1}
\frac{\partial \Tilde{f}}{\partial x}  = \frac{1}{\Delta-\delta} \frac{\partial h(x)}{\partial x}-\mathcal{D}^{in} \left([\text{diag}\left(u(s) \mathds{1}_{s<\alpha(x)}\right)]\otimes I_d \right) \mathcal{D}^T \\
- \mathcal{D}^{in} \text{diag}\big(\mathcal{D}^Tx\big)  \left([u(s)  \text{diag}(\delta(\alpha(x)-s))]\otimes \mathbf{1}_d\right)\frac{\partial \alpha(x)}{\partial x}, 
\end{multline*}
with $\delta(\alpha(x)-s)$ representing the Dirac delta function,
and $\int_{-\infty}^{\infty} \delta(S) ds =1$ and in general case $\int_{-\infty}^{\infty} g(S)\delta(S) ds =g(0)$.
Finally,
\begin{equation*}\label{dfdu for d>1}
\frac{\partial \Tilde{f}}{\partial u(s)} = -D^{in}\text{diag}\big(\mathcal{D}^Tx\big)\left(\mathds{1}_{s<\alpha(x)}\otimes \mathbf{1}_d\right).   
\end{equation*}
Also, we can determine 
$[\frac{\partial \alpha}{\partial x}] = [\frac{\partial \alpha}{\partial z}] [\frac{\partial z}{\partial x}].$

It can  be seen that
\begin{equation}\label{dalphadz}
    [\frac{\partial \alpha_i}{\partial z_j}]_{i, d_j-(d-1):d_j} = \begin{cases}
        \frac{z_i^T}{\|z_i\|} & i=j \\
        0 & i \neq j
    \end{cases}
\end{equation}
So, in compact form
\begin{equation}
 [\frac{\partial \alpha}{\partial z}]  =  \begin{bmatrix}
    \frac{z_{1}^T}{\|z_{11}\|} & 0^T &\dots& 0^T \\
    0^T & \frac{z_2^T}{\|z_{12}\|} &\dots & 0^T \\
    \vdots & & \ddots\\
    0^T &\dots & &\frac{z_{m}^T}{\|z_{md}\|}
\end{bmatrix}_{m \times md} 
\end{equation}
 suggests that $ [\frac{\partial \alpha}{\partial z}]$ is a block diagonal matrix, where each block is a row vector of size $d$. Substituting for $z$, we can see that equation (\ref{dalphadz}) can be written as
\begin{align}
[\frac{\partial \alpha}{\partial x}]_{ki}  &=  \frac{\partial \alpha_k}{\partial x_i} \\ &=
     \begin{cases}
      \frac{(x_i-x_j)^T}{\|x_i-x_j\|} & \textit{if } k \in E, \textit{ where } k=(i,j) || k= (j,i)\\
      0^T & \textit{o.w}
    \end{cases}
\end{align} 

Further, $[\frac{\partial z}{\partial x}] = \mathcal{D}^T$.
So, we can write
$[\frac{\partial \alpha}{\partial x}]^T = \mathcal{D} [\frac{\partial \alpha}{\partial z}]^T, $   
which ties the operation to the graph's topology as follows
\begin{equation}
[\frac{\partial \alpha}{\partial x}]_{kj}  =  \frac{\partial \alpha_k}{\partial x_j}  =
     \begin{cases}
      \frac{(x_i-x_j)^T}{\|x_i-x_j\|} & \textit{if } \exists k \in (j,i)\\
      -\frac{(x_i-x_j)^T}{\|x_i-x_j\|} & \textit{if } \exists k \in (i,j)\\
      0^T & \textit{o.w}
 \end{cases}.
\end{equation}

So, equation (\ref{augmented cost variation 2}) can be simplified to 
\begin{equation}\label{variation of augmented cost-continuous-further simplified 2}
\begin{aligned}
\delta \Tilde{J} &= \frac{\partial \psi}{\partial x}\bigg|_{x=x(t_f)}\eta(t_f)
+\displaystyle\int_{t_0}^{t_f} \int_{\delta}^{\Delta}\frac{\partial\mathcal{I}}{\partial x} \bigg|_{x=x} \eta(\tau)+\frac{\partial\mathcal{I}}{\partial u(s)} \mu(s)\\
& + \lambda^T\Big( \frac{1}{\Delta-\delta}\dot{\eta}(\tau)
 - \frac{\partial \Tilde{f}}{\partial x}\eta(\tau) - \frac{\partial \Tilde{f}}{\partial u(s)} \mu(s)\Big) \, ds d\tau
\end{aligned}
\end{equation}
Using integration by parts and assuming $\eta(t_0)=0$ we can remove the dependency of $\dot{\eta}(\tau)$ as:
\begin{equation}\label{integration by parts}
 \int_{t_0}^{t_f} \lambda^T \dot{\eta}(\tau)d\tau = \lambda^T (t_f) \eta(t_f) -\int_{t_0}^{t_f}\dot{\lambda}^T \eta(\tau)d \tau.     
\end{equation}
So, equation \eqref{variation of augmented cost-continuous-further simplified 2} become:
\begin{equation}\label{using integration by parts}
\begin{aligned}
\delta \Tilde{J} &= \frac{\partial \psi}{\partial x}\bigg|_{x=x(t_f)}\eta(t_f)+\lambda^T(t_f)\eta(t_f)\\
&+\displaystyle\int_{t_0}^{t_f} \int_{\delta}^{\Delta}\frac{\partial\mathcal{I}}{\partial x} \bigg|_{x=x} \eta(\tau)+\frac{\partial\mathcal{I}}{\partial u(s)} \mu(s)\\
& -  \frac{1}{\Delta-\delta}\dot{\lambda}^T\eta(\tau)
 - \lambda^T\Big(\frac{\partial \Tilde{f}}{\partial x}\eta(\tau) + \frac{\partial \Tilde{f}}{\partial u(s)} \mu(s)\Big) \, ds d\tau
\end{aligned}
\end{equation}
Factoring similar terms, we can further simplify the expression for $\delta J$
\begin{equation}\label{final simplification}
\begin{aligned}
\delta \Tilde{J} &= \frac{\partial \psi}{\partial x}\bigg|_{x=x(t_f)}\eta(t_f)+\lambda^T(t_f)\eta(t_f)\\
&+\displaystyle\int_{t_0}^{t_f} \int_{\delta}^{\Delta}\Big(\frac{\partial\mathcal{I}}{\partial x} \bigg|_{x=x}-\frac{1}{\Delta-\delta}\dot{\lambda}^T
 -\lambda^T \frac{\partial \Tilde{f}}{\partial x}\Big)\eta(\tau)\\&+\Big(\frac{\partial\mathcal{I}}{\partial u(s)}  - \lambda^T\frac{\partial \Tilde{f}}{\partial u(s)} \Big)\mu(s) \, ds d\tau,
\end{aligned}
\end{equation}
which can be written as

\begin{equation}
    \begin{aligned}
     \delta \Tilde{J} &= \frac{\partial \psi}{\partial x}\bigg|_{x=x(t_f)}\eta(t_f)+\lambda^T(t_f)\eta(t_f)\\
&+\displaystyle\int_{t_0}^{t_f} \Big[-\dot{\lambda}^T+\int_{\delta}^{\Delta}\Big(\frac{\partial\mathcal{I}}{\partial x} \bigg|_{x=x}
 -\lambda^T \frac{\partial \Tilde{f}}{\partial x}\Big)ds\Big]\eta(\tau)\, d\tau\\&+\int_{t_0}^{t_f} \int_{\delta}^{\Delta}\Big(\frac{\partial\mathcal{I}}{\partial u(s)}  - \lambda^T\frac{\partial \Tilde{f}}{\partial u(s)} \Big)\mu(s) \, ds d\tau,
    \end{aligned}
\end{equation}
Finally, we can set $\delta \Tilde{J}=0$ to formulate the necessary conditions for optimality:
First, given the definition for $\mathcal{I}$ from equation (\ref{definition of I}), the co-state equation becomes:  
\begin{equation}
    \begin{gathered}
        \dot{\lambda}^T  = \int_{\delta}^{\Delta} \Big( \frac{\partial\mathcal{I}}{\partial x} \bigg|_{x=x}  - \lambda^T \frac{\partial \Tilde{f}}{\partial x} \Big)\, ds \\
         = (x-\hat{x})^T-\lambda^T \frac{\partial h(x)}{\partial x} (\Delta-\delta)\\+\lambda^T\int_{\delta}^{\Delta}\mathcal{D}^{in} \big[\text{diag}\left(u(s) \mathds{1}_{s<\alpha(x)}\right)\otimes I_d\big]  \mathcal{D}^T  \,ds\\+ \lambda^T\mathcal{D}^{in} \text{diag}\big(\mathcal{D}^Tx\big)\\  \quad \Big[\int_{\delta}^{\Delta} u(s)  \text{diag}(\delta(\alpha(x)-s))\,ds \otimes \mathbf{1}_d\Big]\frac{\partial \alpha(x)}{\partial x},\\
          = (x-\hat{x})^T-\lambda^T \frac{\partial h(x)}{\partial x}(\Delta-\delta)+\lambda^T\mathcal{D}^{in} \mathbb{W}  \mathcal{D}^T\\+ \lambda^T\mathcal{D}^{in} \text{diag}\big(\mathcal{D}^Tx\big) \\  \quad \Big[\int_{\delta}^{\Delta} u(s)  \text{diag}(\delta(\alpha(x)-s))\,ds \otimes \mathbf{1}_d\Big]\frac{\partial \alpha(x)}{\partial x}\\
         =(x-\hat{x})^T-\lambda^T \frac{\partial h(x)}{\partial x}(\Delta-\delta)+\lambda^T\mathcal{D}^{in} \mathbb{W}  \mathcal{D}^T \\+ \lambda^T\mathcal{D}^{in} \text{diag}\big(\mathcal{D}^Tx\big)  \big[\text{diag}\left(u(\alpha(x)\right) \otimes \mathbf{1}_d\big]\frac{\partial \alpha(x)}{\partial x}\\
          =(x-\hat{x})^T-\lambda^T \frac{\partial h(x)}{\partial x}(\Delta-\delta)+\lambda^T\mathcal{D}^{in} \mathbb{W}  \mathcal{D}^T \\+ \lambda^T\mathcal{D}^{in} \odot x^T\mathcal{D}  \big[\text{diag}\left(u(\alpha(x)\right) \otimes \mathbf{1}_d\big]\frac{\partial \alpha(x)}{\partial x}.
    \end{gathered}
\end{equation}
Note that $\text{diag}(\mathcal{D}^Tx)=\text{diag}(x^T\mathcal{D})$.

The boundary condition is:
\begin{equation}
 \lambda (t_f)=-\Big(\frac{\partial \psi}{\partial x}\bigg|_{x=x(t_f)}\Big)^T.
\end{equation}
Finally, we set the last term of equation (\ref{final simplification}) to zero. We can see that $\mu(s)$ does not depend on time, so we can swap the integrals as follows

\begin{equation}
    \begin{aligned}
        \int_{\delta}^{\Delta} \Big[\int_{t_0}^{t_f} \Big(\frac{\partial\mathcal{I}}{\partial u(s)}  - \lambda^T\frac{\partial \Tilde{f}}{\partial u(s)} \Big) \, d\tau\Big]\,\mu(s) ds  = 0.
    \end{aligned}
\end{equation}
We have an additional integral constraint, giving us the optimal policy
\begin{equation}
    \begin{aligned}
     \int_{t_0}^{t_f}\frac{\partial\mathcal{I}}{\partial u(s)}  - \lambda^T\frac{\partial \Tilde{f}}{\partial u(s)}   \,d\tau =0,
    \end{aligned}
\end{equation}
which results on a condition on the optimal policy, $u(s)$ as follows
\begin{multline}\label{Ju}
     \Tilde{J}_u= u-u_o+ \tfrac{1}{t_f-t_0}\int_{t_0}^{t_f}\lambda^T \mathcal{D}^{in}\text{diag}(\mathcal{D}^Tx)[\mathds{1}_{s<\alpha(x)}\otimes \mathbf{1}_d] \, d\tau\\
     = u-u_o+ \tfrac{1}{t_f-t_0}\int_{t_0}^{t_f} \sum_{i=1}^N\sum_{j\in \mathcal{N}_i^{in}}\lambda_i^T (x_i-x_j)\mathds{1}_{s<\|x_i-x_j\|} \, d\tau
\end{multline}

\subsection{Hessian Calculation for Second-order Condition}

To verify the second-order sufficient condition, we examine the second derivative of the Hamiltonian w.r.t the control input $ u(s) $. Specifically, the Hessian of the Hamiltonian is:
\[
\frac{\partial^2 H}{\partial u^2}(x^*, u^*, \lambda^*) = \int_{\delta}^{\Delta} \frac{\partial^2 I(x^*, u^*)}{\partial u^2} + \lambda^T \frac{\partial^2 \tilde{f}(x^*, u^*)}{\partial u^2} \, ds.
\]
Since $\mathcal{I}(x, u(s)) $ is quadratic in $ u(s) $ (due to the regularization term $ G(u(s)) = \frac{1}{2} \| u(s) - u_o(s) \|^2 $), the second derivative $ \frac{\partial^2 \mathcal{I}(x, u(s))}{\partial u(s)^2} $ is positive definite. This ensures that the contribution from $ \mathcal{I}(x, u(s)) $ to the Hessian is strictly positive.

For the dynamics $ \tilde{f}(x, u(s)) $, the term $ \frac{\partial^2 \tilde{f}(x, u(s))}{\partial u(s)^2} $ is zero because $ \tilde{f}(x, u(s)) $ is linear in $ u(s) $. Therefore, the Hessian simplifies to:
\[
\frac{\partial^2 H}{\partial u^2}(x^*, u^*(s), \lambda^*) = \int_{\delta}^{\Delta} \frac{\partial^2 I(x^*, u^*(s))}{\partial u^2(s)} \, ds.
\]
Since $ \frac{\partial^2 I(x^*, u^*(s))}{\partial u^2(s)} > 0 $ for all $ s \in [\delta, \Delta] $, the Hessian is strictly positive definite. Therefore, the second-order sufficient condition for optimality is satisfied, ensuring that the solution $ u^*(s) $ corresponds to a strict minimum of the cost functional. Thus, the control problem defined by \eqref{cost} satisfies the second-order sufficient condition for optimality.


\addtolength{\textheight}{12cm}   

\bibliographystyle{IEEEtran}
\bibliography{references}

\end{document}